\documentclass[aps,prd,twocolumn,10pt,amsmath,reprint,groupedaddress,amssymb,amsfonts,nofootinbib,longbibliography]{revtex4-1}

\usepackage{graphicx}
\usepackage{amssymb}
\usepackage{amsmath}
\usepackage{longtable}
\usepackage[utf8]{inputenc} 
\usepackage{color}
\usepackage{supertabular}
\usepackage{dcolumn}
\usepackage{ wasysym }
\usepackage{subfigure}
\usepackage{notoccite}

\def\Tcoh{T_{\textrm{\mbox{\tiny{coh}}}}}

\def\EatH{Einstein@Home}

\def\CR{\textrm{CR}}
\def\SDd{{\mathcal{D}}}

\def\totalTemplates{4.99\times 10^{17}}

\newcommand{\avgSeg}[1]{\overline{#1}}			

\newcommand{\Freq}{f}
\newcommand{\fdot}{{\dot{\Freq}}}
\newcommand{\fddot}{\ddot{\Freq}}

\newcommand{\Gauss}{\mathrm{\MakeUppercase{G}}}
\newcommand{\Signal}{{\mathrm{\MakeUppercase{S}}}}
\newcommand{\Line}{{\mathrm{\MakeUppercase{L}}}}
\newcommand{\Noise}{{\Gauss\Line}}

\providecommand{\sc}[1]{\widehat{#1}}
\renewcommand{\sc}[1]{\widehat{#1}}

\newcommand{\OSNsc}{\sc{O}_{{\Signal\Noise}}}	

\newcommand{\F}{\mathcal{F}}		

\newcommand{\avF}{\avgSeg{\F}}

\newcommand{\Nseg}{{N_{\mathrm{seg}}}}

\let\svthefootnote\thefootnote

\begin{document}
\date{\today}
\title{ An Einstein@Home search for continuous gravitational waves from Cassiopeia A }
\author{Sylvia J. Zhu$^\mathrm{1,2,a}$, Maria Alessandra Papa$^\mathrm{1,2,4,b}$, Heinz-Bernd Eggenstein$^\mathrm{2,3}$, Reinhard Prix$^\mathrm{2,3}$, Karl Wette$^\mathrm{2,3}$, Bruce Allen$^\mathrm{2,4,3}$, Oliver Bock$^\mathrm{2,3}$, David Keitel$^\mathrm{2,3,5}$, Badri Krishnan$^\mathrm{2,3}$, Bernd Machenschalk$^\mathrm{2,3}$, Miroslav Shaltev$^\mathrm{2,3}$, Xavier Siemens$^\mathrm{4}$\\\vspace{0.3in}
}\let\thefootnote\relax\footnote{\textsuperscript{a}email: sylvia.zhu@aei.mpg.de}\let\thefootnote\relax\footnote{\textsuperscript{b}email: maria.alessandra.papa@aei.mpg.de}
\affiliation{$^1$ Max-Planck-Institut f{\"u}r Gravitationsphysik, am M{\"u}hlenberg 1, 14476, Potsdam-Golm\\
$^2$ Max-Planck-Institut f{\"u}r Gravitationsphysik, Callinstra{$\beta$}e 38, 30167, Hannover\\
$^3$ Leibniz Universit{\"a}t Hannover, Welfengarten 1, 30167, Hannover\\
$^4$ University of Wisconsin-Milwaukee, Milwaukee, Wisconsin 53201, USA\\
$^5$ Universitat de les Illes Balears, IAC3---IEEC, E-07122 Palma de Mallorca, Spain\\
\vspace{0.1in}}
\addtocounter{footnote}{-2}\let\thefootnote\svthefootnote

\begin{abstract}
  We report the results of a directed search for continuous gravitational-wave
  emission in a broad frequency range (between 50 and 1000 Hz) 
  from the central compact object of the supernova remnant
  Cassiopeia A (Cas A). The data comes from the sixth science run of LIGO and the search 
  is performed on the volunteer distributed computing
  network Einstein@Home. We find no
  significant signal candidate, and set the most constraining upper limits to date
  on the gravitational-wave emission from Cas A, which beat the indirect age-based upper limit across the 
  entire search range. At 170~Hz 
  (the most sensitive frequency range), we set $90\%$ confidence upper limits
  on the gravitational wave amplitude $h_0$ of
  $\sim\!\!~2.9\times 10^{-25}$, roughly twice as constraining
  as the upper limits from previous searches on Cas A. The upper limits can also be expressed as constraints
  on the ellipticity of Cas A; with a few reasonable assumptions, we show that
  at gravitational-wave frequencies greater than 300~Hz, we can exclude an ellipticity
  of $\gtrsim\!\!~10^{-5}$.
\end{abstract}

\pacs{}
\preprint{LIGO-P}
\maketitle

\section{Introduction}
\label{sec:introduction}

Isolated neutron stars with non-axisymmetric asymmetries are thought to be one of the best
sources for continuous gravitational-wave emission. We report the results of a
directed search for continuous gravitational-wave emission
from the central compact object of the supernova remnant Cassiopeia A (Cas A) with the
Laser Interferometer Gravitational-Wave Observatory (LIGO). Directed searches, in which
the source and therefore the sky position are specified, are generally more sensitive than
all-sky surveys. The reason is that typically fewer templates are needed for directed
searches than for all-sky surveys; this results in a smaller trials factor and hence
in a smaller weakest detectable signal at fixed detection confidence.



At an age of a few hundreds of years, Cas A is one of the youngest known supernova
remnants \cite{CasA:age}. Its young age means that any asymmetries in the central compact object that
were produced at birth are likely still present. Based on X-ray observations, the
central compact object is most likely a neutron star with a low surface
magnetic field strength \cite{CasA:NS}. No pulsed electromagnetic
emission has been observed from the central object, so its spin parameters are unknown.

Assuming the central object is a neutron star, its asymmetries would be
expected to continuously produce slowly evolving and nearly monochromatic gravitational waves
(e.g., \cite{S5CasA}). We perform a
search for this gravitational-wave emission from Cas A using data from the sixth
LIGO science run with the volunteer distributed computer
network Einstein@Home \cite{EatH}. 

For the remainder of this paper, when we refer to Cas A, we are referring to the
central compact object.

\section{The search}

\subsection{Data used in this search}

The two LIGO interferometers are located in the US in Hanford, Washington and Livingston, Louisiana,
a separation distance of 3000~km \cite{LIGO:detector}. The last science run of initial LIGO, S6,
took place between July 2009 and October 2010 \cite{LIGO:S6}. For this analysis, we only use
data taken between 06 February 2010 (GPS 949461068) and 20 October 2010 (GPS 971629632),
selected for the best sensitivity \cite{Shaltev:2015saa}.

Unlike what we did for previous Einstein@Home searches \cite{S6Bucket,S5GC1HF},
we do not perform any upfront line cleaning to remove known artefacts.

\subsection{The search set-up}

\begin{table}
  \centering
  \begin{tabular}{|c|c|}
    \hline \hline
    $T_\textup{coh}$ (hrs)            & 140 \\ \hline
    $t_\textup{ref}$ (GPS s)          & 960541454.5 \\ \hline
    $N_\textup{seg}$                  & 44  \\ \hline
    $\delta f$ (Hz)                  & $5.4\times 10^{-7}$ \\ \hline
    $\delta \dot{f}$ (Hz\,s$^{-1}$)   & $8.2 \times 10^{-12}$ \\ \hline
    $\delta \ddot{f}$ (Hz\,s$^{-2}$)  & $1.9 \times 10^{-18}$ \\ \hline
    $\gamma_1$                       & 90  \\ \hline
    $\gamma_2$                       & 60 \\
  \hline \hline
  \end{tabular}
  \caption{The search parameters (rounded to the first decimal point) are
    listed. $t_\textup{ref}$ is the reference time at which
    the values of $f$ and $\dot{f}$ are defined. $\gamma_1$ and $\gamma_2$
    are the refinement factors for the $\dot{f}$ and $\ddot{f}$
    grids, respectively, during the incoherent summation stage.}
  \label{parameters}
\end{table}

\begin{figure}[h]
  \includegraphics[width=\columnwidth]{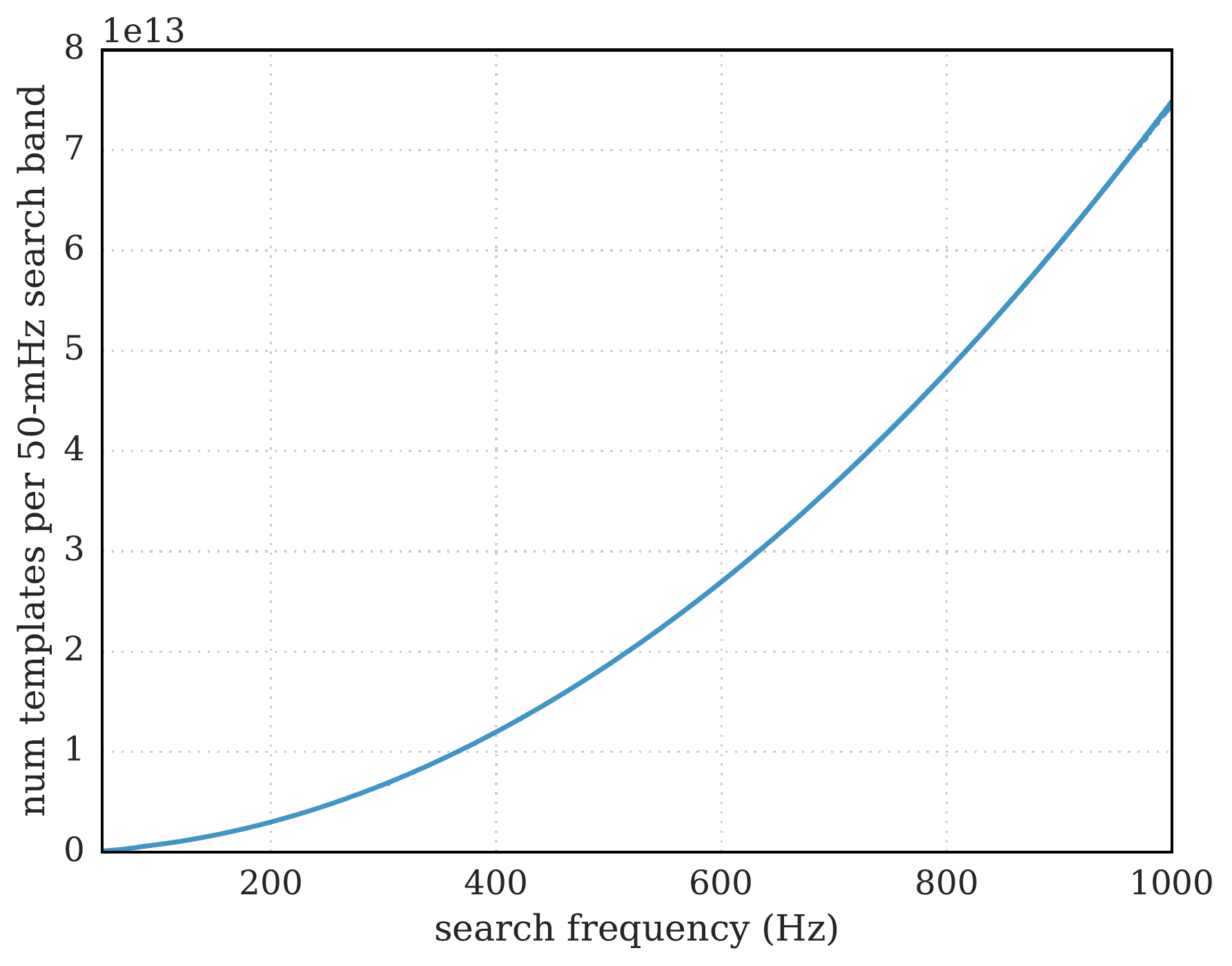}
   \caption{For this search, the number of templates per 50-mHz search band increases
     quadratically with $\Freq$. At each value of $\Freq$, the $\fdot$ search
     range is [-$\Freq/\tau_\textrm{NS}$, 0] and the $\fddot$ search range is
     [0, $2\Freq/\tau_\textrm{NS}^2$]. In total, $\totalTemplates$ templates are
     included in this search.}
\label{fig:templateCounts}
\end{figure}

We perform a semi-coherent search
and rank the results according to the line-robust statistic $\OSNsc$ \cite{Keitel2014},
in a manner similar to \cite{S6Bucket, S5GC1HF}.
The basic ingredient is the averaged $\F$ statistic \cite{JKS, Cutler+Schutz}, $\avF$,
computed using the Global Correlation Transform (GCT) method \cite{Pletsch+Allen,Pletsch2010}.
In a stack-slide search, the time series data are partitioned into $i=1...\Nseg$ segments
of length $T_\mathrm{coh}$ each. The data from every segment are match-filtered against a set of signal
templates each specified by a set of parameters (the signal frequency $\Freq$,
the first-order spindown $\fdot$, the second-order spindown $\fddot$, and the sky position)
to produce values of the detection statistic $\F_i$ for each segment (the coherent step). These
$\F_i$ values are then combined to produce an average value of the statistic across the
$\Nseg$ segments, $2\avF$, which is the core statistic that we use in these analyses (the incoherent step). 
%
The values of the signal template parameters $\Freq$, $\fdot$,
  and $\fddot$ are given by a predetermined grid.
The $\Freq$ grid spacing (i.e., the separation between
two adjacent values of $\Freq$ in the search) is kept the same for the
coherent and incoherent steps, while the spacings for the $\fdot$ and
$\fddot$ grids for the incoherent summing are finer by factors of 90 and 60, respectively.
The search parameters are summarized in Table~\ref{parameters} and were derived 
using the optimisation scheme described in \cite{Prix:2012yu} assuming a run duration of 6 months on \EatH.

\subsection{The detection and ranking statistics}
\label{statistics}

The 2$\avF$ statistic gives a measure of the likelihood that the data
resembles a signal versus Gaussian noise; therefore, signals
are expected to have high values of $2\avF$.
However, line disturbances in the data can also result in high values
of $2\avF$. The line-robust statistic, $\OSNsc$, was designed to address
this by testing the signal hypothesis against a composite noise model comprising 
a combination of Gaussian noise and a  single-detector spectral line. The $\OSNsc$ parameters are
tuned as described in \cite{Keitel2014} using simulations so that the
detection efficiency of $\OSNsc$ performs as well as $2\avF$ in Gaussian
noise and better in the presence of lines. For this search, the value of $c^\star$
(related to the tuning parameter in the choice of prior, see \cite{Keitel2014}) is set to be $34.8$,
which corresponds to a Gaussian false-alarm probability  of $10^{-9}$.

The $\OSNsc$ distribution even in Gaussian noise is not known analytically. Therefore,
although we use the $\OSNsc$ toplists to find the best signal candidates,
we still use $2\avF$ as the detection statistic for ascertaining a candidate's
significance. For a stack-slide search with $\Nseg$ segments,  the $\Nseg \times 2\avF$
distribution in Gaussian noise follows a chi-squared distribution
with $4\Nseg$ degrees of freedom \cite{Wette2012}.

\subsection{The parameter space}

Since the spin parameters of Cas A are unknown, our search encompasses
a large range of possible gravitational-wave frequencies $\Freq$;
namely, from 50 to 1000 Hz. For a given value of $\Freq$,
the $\fdot$ and $\fddot$ ranges are given by the following specifications:
\begin{align}
  \fdot  & \in [-\Freq/\tau_\mathrm{NS}, 0] \\
  \fddot & \in [0, 2\Freq/\tau_\mathrm{NS}^2] 
  \label{eq:f-fdotBox}
\end{align}
where $\tau_\mathrm{NS}$ is the fiducial age of the neutron star, taken to
be 300 years. As discussed in \cite{S5CasA}, this choice of $\tau_\mathrm{NS}$
is on the young end of the age estimates, which yields a larger search parameter space
than other, less conservative choices. The searched parameter space at each value of $\Freq$
is a rectangle in the $\fdot - \ddot{f}$ plane, and the search volume increases quadratically with $\Freq$
(Fig.~\ref{fig:templateCounts}).

Compared to \cite{Wette:2008hg, S5CasA, NineYoungSNRs}, the largest magnitude
of the first order spindown parameter is the same, corresponding to a conservative
assumption (in the sense that it allows for the broadest range of first order
spindown values) on the average braking index at fixed age of the object.
The range of the second order spindown is constructed differently here than
in \cite{S5CasA, NineYoungSNRs} in that it does not depend on $\fdot$.
The highest searched value of $\ddot{f}$ is $n {{{\fdot^2}_{\textrm{max}}} / f}$, 
with $n=2$ being the instantaneous braking index. The searches mentioned
above took this as the {\it lower} boundary of the $\ddot{f}$ range and
set the upper boundary at $n=7$. Our choice does not search such a broad
range of $\ddot{f}$ values, and is driven by ease of set-up of the search.
Observational data on braking indexes supports this choice \cite{Hamil:2015hqa}.

We estimate that searching over third order spindown values is not necessary.
We do this by counting how many templates are needed to cover the third order
spindown range. The 3rd order spindown template extent ${\Delta\dddot{f}}$ in a semi-coherent search with mismatch $m$ is
\begin{equation}
{\Delta}\dddot {f} ={1\over \gamma_3}{{2520\sqrt{m}}\over{\pi \Tcoh^4}}.
\label{eq:extensionThirdOrderSpindown}
\end{equation}
For this search we set $m=0.2$ and $\gamma_3\simeq3.89\times 10^5$ \cite{ShaltevGammaRef:2012}. The template extent of
Eq.~\ref{eq:extensionThirdOrderSpindown} is $\sqrt{m g^{33}}$
where $g^{ij}$ is the inverse of the phase metric \cite{Wette:2015lfa}.
The 3rd order spindown range, consistent with
the choices of Eq.~\ref{eq:f-fdotBox}, is $6 {f / \tau_{\textrm{NS}}^3}$. With $\tau_\mathrm{NS} = 300$~years,
we find that we do not need more than a single template to cover the 3rd order spindown range;
therefore, we do not need to add a 3rd order spindown dimension.

The location of Cas A is known to within $\sim$ 1", which is smaller than the
sky resolution of our search. Hence, we only search a single sky position
(right ascension = 23h~23m~28s, declination = 58$^\circ$~58'~43'').


\subsection{Distribution of the computational load}

The search runs on volunteer computers in the Einstein@Home network, and is
split into 9.2~million work units (WUs), with each WU designed to run for about 
six hours on a modern PC. A single WU encompasses a 50-mHz
range in $\Freq$ and the entire range of $\fddot$ at the start value of $\Freq$,
along a single slice out of the $\fdot$ range. The results from WUs that search
over the same 50-mHz range are combined into a single band, and these multiple WUs together cover
the entire $\fdot$ range at that value of $\Freq$.
Each WU searches through
approximately $5 \times 10^{10}$ templates, and returns
two lists of results corresponding to the 3000 templates with the highest values of the 2$\F$ and $\OSNsc$
statistics (described in section~\ref{statistics}), called the toplists.
The total number of templates included in this search is $\totalTemplates$.

\subsection{Semiautomatic identification of disturbances}

\begin{figure*}
  \includegraphics[width=0.8\textwidth]{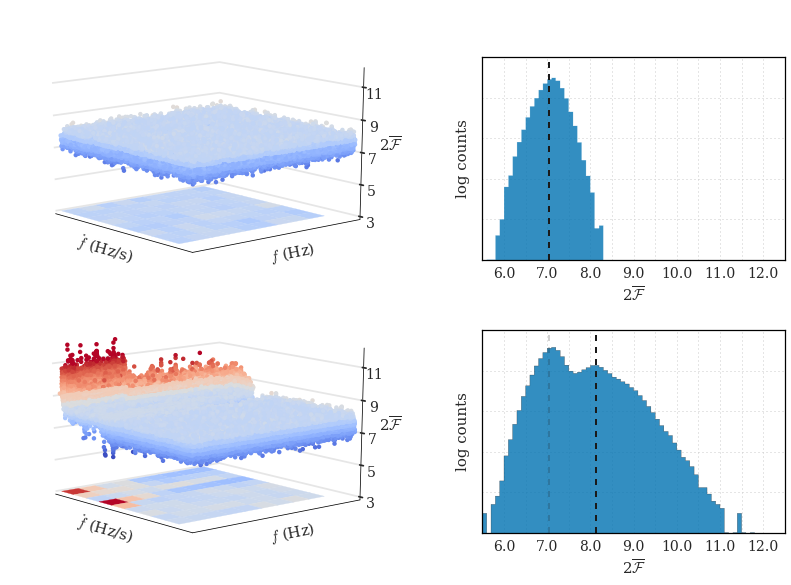}
  \caption{The maximum density (left) and the mean $2\avF$ value (right) for the
    candidates are the two metrics we use to identify potentially disturbed
    bands. \textbf{Left:} undisturbed bands (an example in the top panels) have a very uniform density
    of candidates in $\Freq$-$\fdot$, while disturbed bands (an example in the bottom panels)
    present marked overdensities. The $2\avF$ values in $\Freq$-$\fdot$ plane are shown in the 3D plot, while
    the candidate density is shown in the 2D projections.The maximum density in
    a disturbed band tends to be much higher (here, more red) than the maximum
    density in an undisturbed band. \textbf{Right:} The $2\avF$ distribution
    in an undisturbed band (top) and in a
    disturbed band (bottom).}
  \label{UvsD}

  \includegraphics[width=0.6\textwidth]{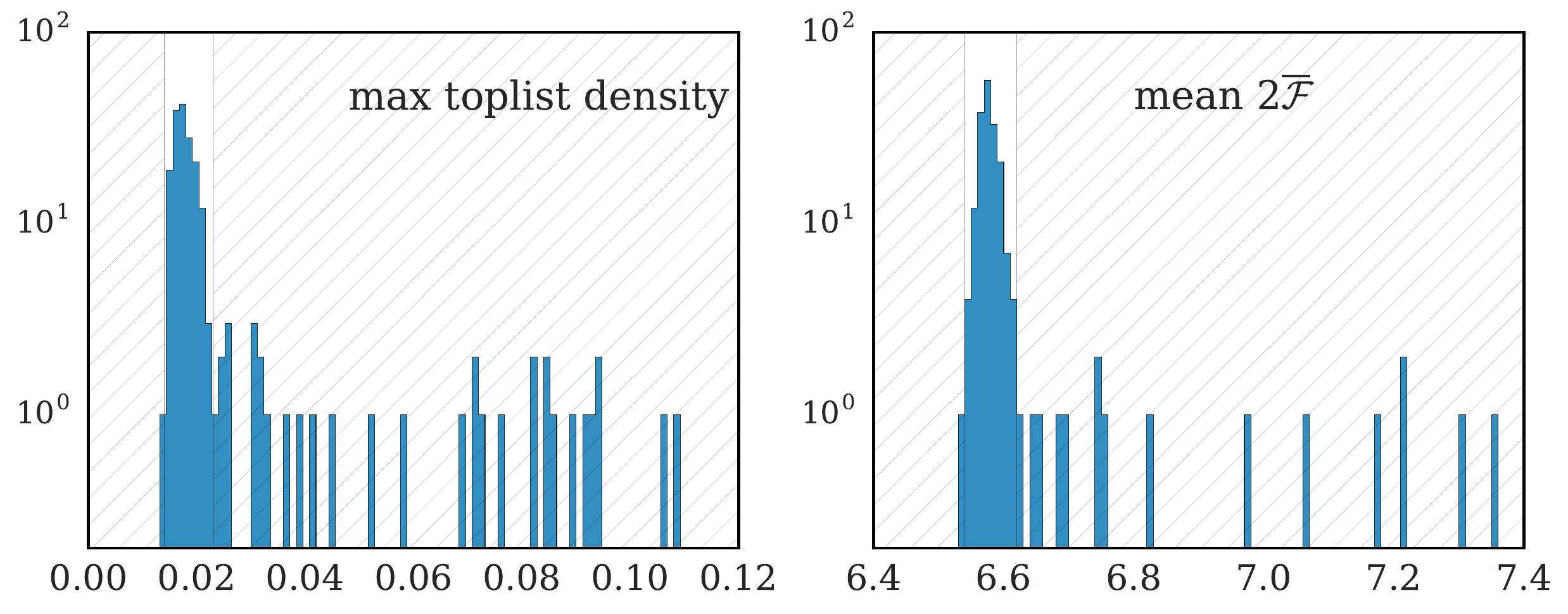}
  \caption{The distributions of maximum toplist density (left) and
  mean toplist $2\avF$ (right) are shown for a sample 10-Hz frequency range.
  Both distributions consist of an undisturbed body with a disturbed tail
  (hatched). All 50-mHz bands that fall within the hatched areas are
  marked as potentially disturbed.}
  \label{fig:systDistId}
\end{figure*}

When the noise is purely Gaussian, the 2$\avF$ distribution is well-modelled
and the significances of signal candidates can be determined in a
straight-forward manner. However, disturbances generate deviations from the expected distribution. 
In order to meaningfully use the same statistical analysis
on all of the candidates, the disturbed 50-mHz bands must be excluded from the
search. Previous searches \cite{S5GC1HF,S6Bucket} relied on a visual inspection of the full
data set in order to identify the disturbed bands, which is a very time
consuming endeavor. Here, we introduce a semiautomatic method that greatly
reduces the number of bands that need to be visually inspected.

We use two indices to identify bands that cannot automatically be
classified as undisturbed: 1) the density of toplist candidates in that band and
2) their average $2\avF$.
We classify as undisturbed those bands whose maximum density and
average $2\avF$ are well within the bulk distribution of the values
for these quantities in the neighbouring frequency bands, and mark
the remainder as potentially disturbed and in need of visual inspection.


The size of the toplist and the frequency grid spacing are fixed. 
Therefore, when a disturbance is present in a 50-mHz band,
the toplists within that band disproportionately include templates
in the parameter space near the disturbance. We look for evidence of
disturbances in the 50-mHz bands using a method that
mimics and replaces the visual inspection used in previous searches
\cite{S6Bucket}: For a given band, we calculate
the density of candidates in a $10 \times 10$ grid in $\Freq$-$\fdot$ space,
and take the maximum density as an indicator of how disturbed
the band is likely to be. Since disturbances also manifest as
deviations in the $2\avF$ distribution, we use the mean of $2\avF$
as an additional indicator of how much a band is disturbed. A visual representation
of these concepts is shown in Fig.~\ref{UvsD}.

Because the search volume increases with $\Freq$, both the 
mean of $2\avF$ and the candidate density vary with frequency. 
To account for this effect, we compare the observed maximum density
and mean $2\avF$ from each band with the distribution of maximum density
and mean $2\avF$ values in sets of 200 contiguous 50-mHz bands (10~Hz).
These constitute our reference distributions.

Since the majority of bands are undisturbed, the reference distributions are
composed of a well-defined bulk (from the undisturbed bands) with
tails (disturbed bands), as illustrated in Fig.~\ref{fig:systDistId}. We
define the ``bulk" of each distribution by eye, and then mark the bands
that fall outside of this bulk on either side as being potentially disturbed;
we generally expect disturbed bands to be in the upper ends of the distributions
(that is, to have particularly large values of maximum density and mean $2\avF$)
but also include bands in the lower ends so as not to miss any unexpected
disturbed behavior. We proceed with a full visual inspection only of
this potentially disturbed subset. 
Fig.~\ref{fig:systDistId} shows the reference distributions for the
bands between 90 and 100~Hz. These are typical examples and illustrate
how the ``by eye'' definition of the bulk of the distributions is
not subtle. When selecting the bulk, we err on being conservative:
when in doubt, we label bands as being potentially disturbed, as
these will be re-inspected later. 

If a signal were present, it would not be excluded because of the
automated procedure. On the one hand, if it were so weak that the band
would not be marked as disturbed, the band would automatically be
included in the analysis. On the other hand, if it were strong
enough that the automated procedure mark its band as being potentially
disturbed, then it would be visually inspected by a human who
would recognise the signature of a signal and not discard the band. 

This method still requires human input in two steps: first, to define
the bulk of the reference distributions; and second, to inspect the
subset of potentially disturbed bands.
However, the ``calibration"
work necessary for determining the bulk of the reference distributions 
only requires the inspection of 2 distributions every 10 Hz rather than multiple
distributions every 50 mHz. Furthermore, the bands that do not pass the
undisturbed-classification criteria and require visual inspection are only 15\%
of the total set. Overall, this procedure still cuts down the required time from
multiple days with multiple people to a few hours by a single person.



This procedure requires minimal tuning and relies only on the assumption
that the reference distributions are predominantly undisturbed. This
has so far been our experience on all the LIGO data sets that we have inspected. We are confident
that this method can be applied to other sets of gravitational wave data. 

When we compare this method against a full
visual inspection of a few search frequency ranges
(50 to 100 Hz, 450 to 500 Hz, and 950 to 1000 Hz), it identifies $\sim\!95\%$ of the disturbed bands
and misses only the most marginal disturbances. After we apply
this method to the entire frequency range, we exclude a total
of 1991 50-mHz bands as being disturbed ($\sim\!10\%$);
these are listed in Table~S2.

\subsection{Analysis of undisturbed bands}

\begin{figure*}
  \includegraphics[width=0.6\textwidth]{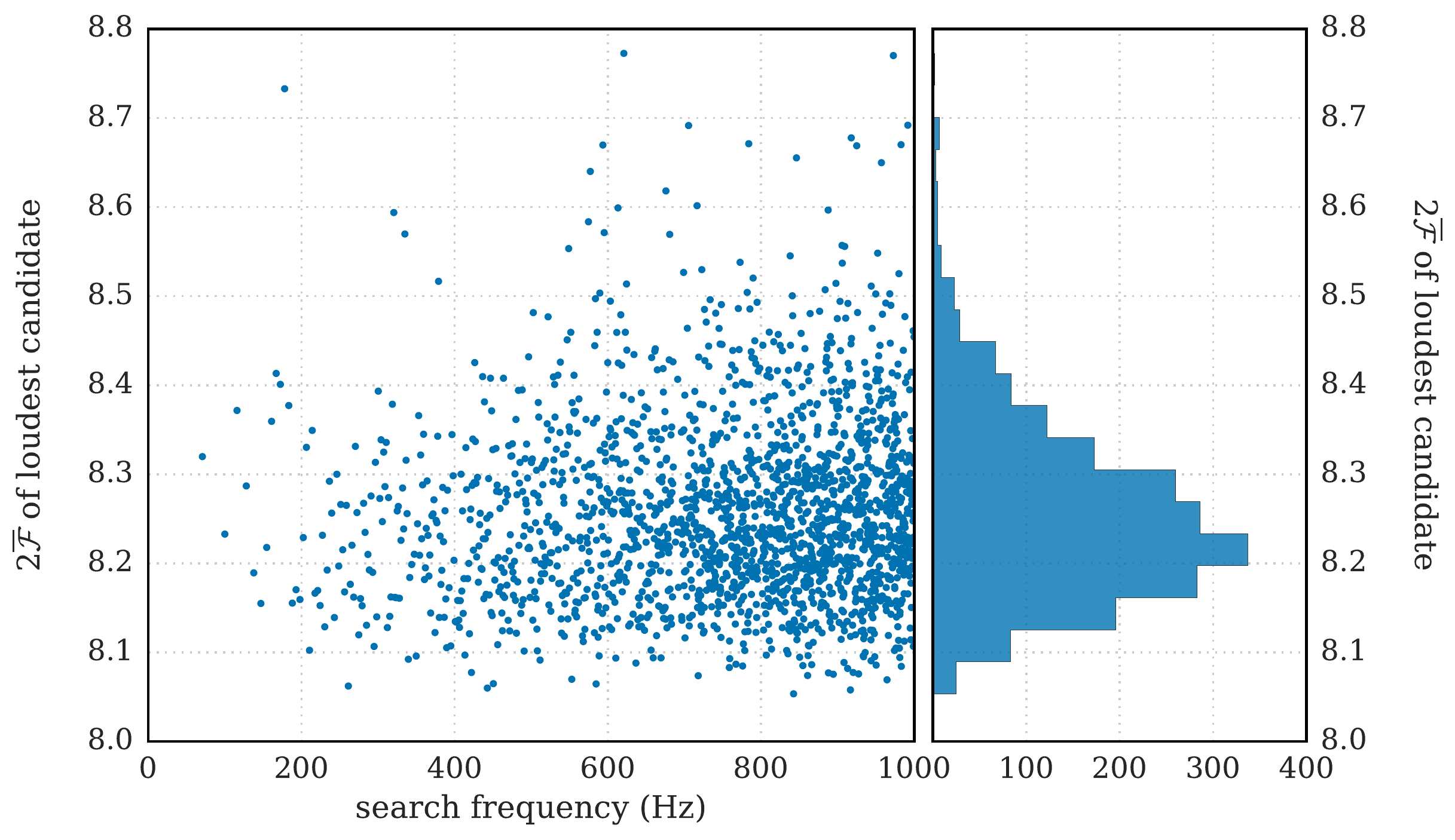}
  \includegraphics[width=0.6\textwidth]{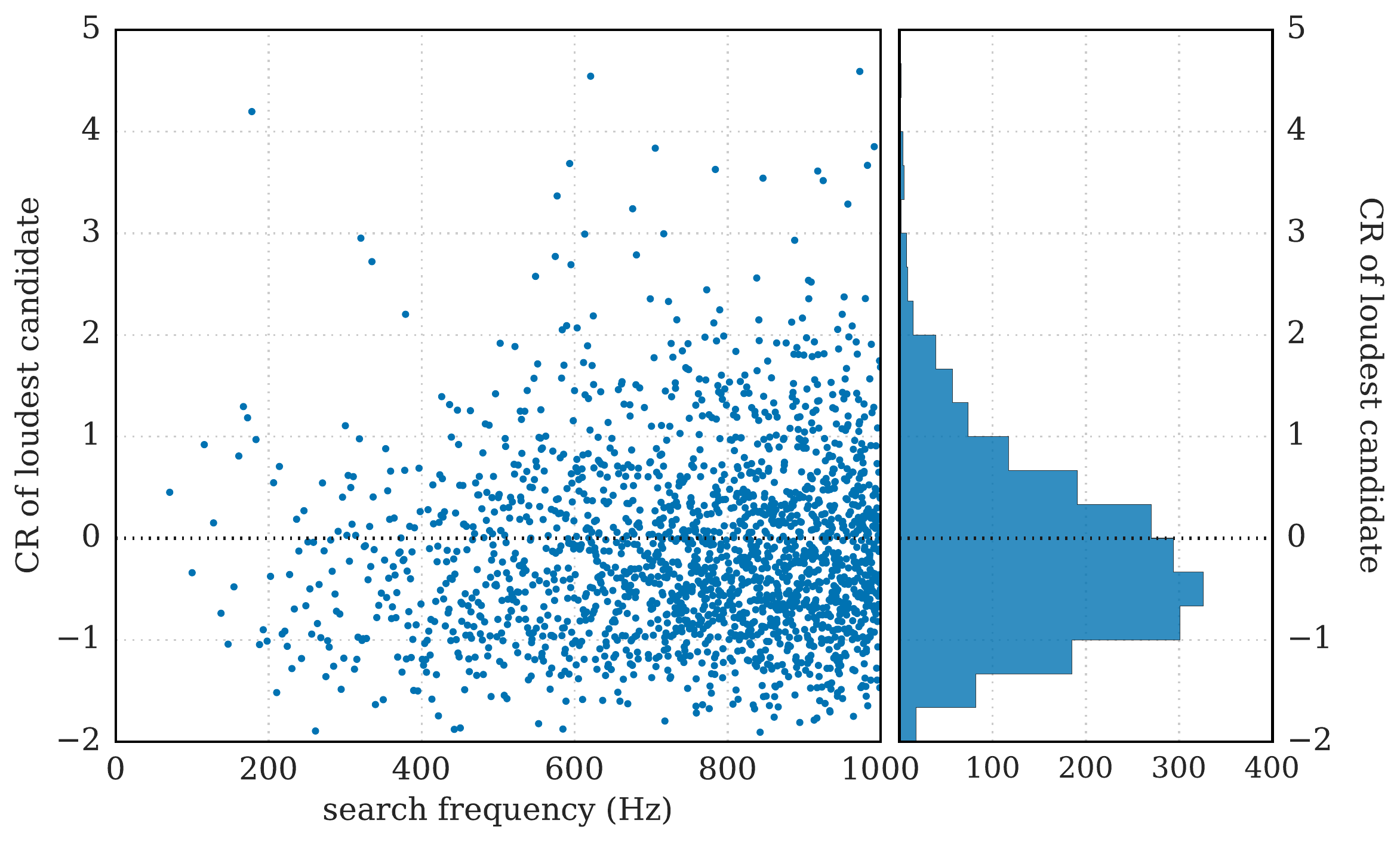}
  \caption{For each of the 2000 partitions, we determine
    the $2\avF$ of the loudest candidates (top) as well as their
    CR values (bottom), where CR is defined in Eq.~\ref{CR}.}
  \label{fig:loudest}
\end{figure*}

The $2\avF$ distribution in Gaussian noise only depends on
the number of effectively independent templates searched ($N$). However, the grid spacings
are chosen to maximize signal recovery, so the $N$ templates
are not fully independent. The observed distribution is
instead described by an effective number of templates
$N_\mathrm{eff} < N$. The value of $N_\mathrm{eff}$ is obtained
by fitting the distribution of the loudest candidates
(i.e., the highest values of $2\avF$).

We divide the entire set of 50-mHz bands across our search frequency
range into 2000 partitions of approximately equal parameter space volume,
which results in $\sim\!2\times10^{14}$ templates per partition. In
order to create these partitions, we calculate an exact partitioning
of the total search volume and divide the full range of 50-mHz bands so that
the number of templates in each partition best matches the number
of templates in the exact partitioning.
Since the number of templates in a band grows with frequency,
the frequency width spanned by each partition decreases with increasing
frequency. This can be clearly seen in the left panels of Fig.~\ref{fig:loudest}. In this way, since each
partition contains roughly the same number of search templates, the
expected loudest candidate in each partition is the same and drawn from the same
underlying distribution, defined by $N_\mathrm{eff}$. For this search,
we find that $N_\mathrm{eff}/N~\approx~0.65$. 

\begin{figure}
  \includegraphics[width=\columnwidth]{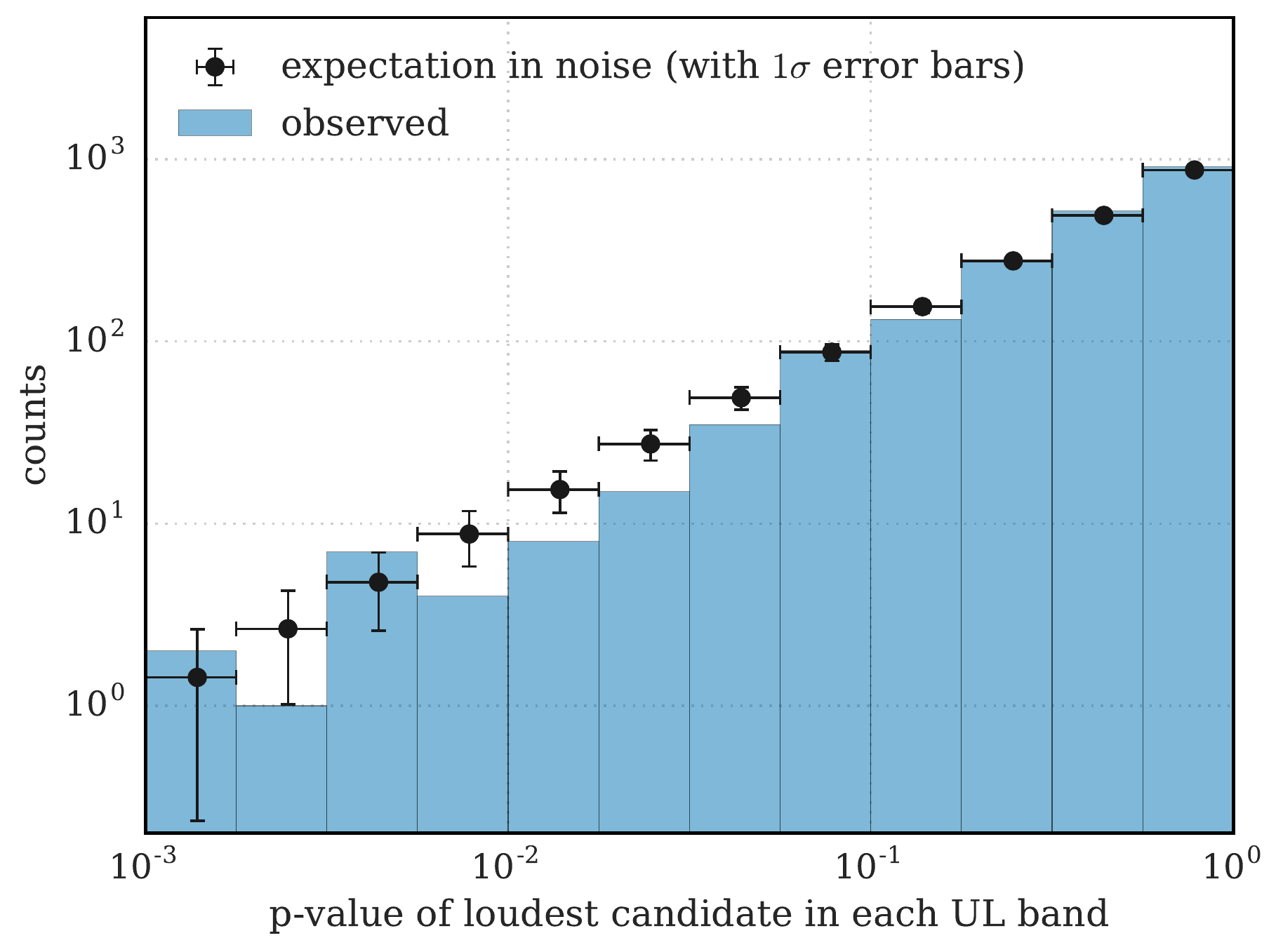}
  \caption{The p-values for the loudest candidate in each UL band
    is plotted in the blue histogram, and the expectation in Gaussian
    noise is shown in the black scatter points. We do not find any
    excess in our search. The small systematic deviation in our
    data from the expected is caused by a subtle difference in
    the $\OSNsc$- and $2\avF$- rankings.}
  \label{pvalues}
\end{figure}

Fig.~\ref{fig:loudest} shows both the $2\avF$ (top) and the
critical ratio CR (bottom) for the loudest candidates. We define CR as
\begin{align}
  CR := \frac{2\avF_\mathrm{meas} - 2\avF_\mathrm{exp}}{\sigma_{2\avF}}
  \label{CR}
\end{align}
where $2\avF_\mathrm{meas}$ is the measured value of the loudest, 
$2\avF_\mathrm{exp}$ the expected value of the loudest, and $\sigma_{2\avF}$
is the  expected standard deviation for the loudest over a partition.
The loudest candidate over the entire search
is in the 620.85~Hz band and has a $2\avF$ value of 8.77;
this is also the most significant candidate, with a CR of 4.56.
However, if we consider the entire searched parameter space rather
than just the partition at 620.85 Hz, the CR value of the most
significant candidate drops to $<0$; i.e., the expected loudest
is actually higher than the loudest that we observe. 
This tells us that our search has not revealed any gravitational wave signal from Cas A in the 
targeted waveform parameter space, as even the template that most resembles a signal has a
statistical significance that is well within the expectations due to random chance.

We convert the CR values of the loudest candidates to p-values to represent the
chance probability of finding a partition-loudest candidate as significant as or
more significant than what was measured in the search.
The results are plotted in Fig.~\ref{pvalues}, along with the expected distribution of
p-values in Gaussian noise.
There is a small systematic deviation
from the expected distribution which arises from a subtle
difference between the $\OSNsc$ and $2\avF$ toplists and is not due to any physical effect.

\section{Upper limits}

\begin{figure*}[ht]
  \includegraphics[width=0.8\textwidth]{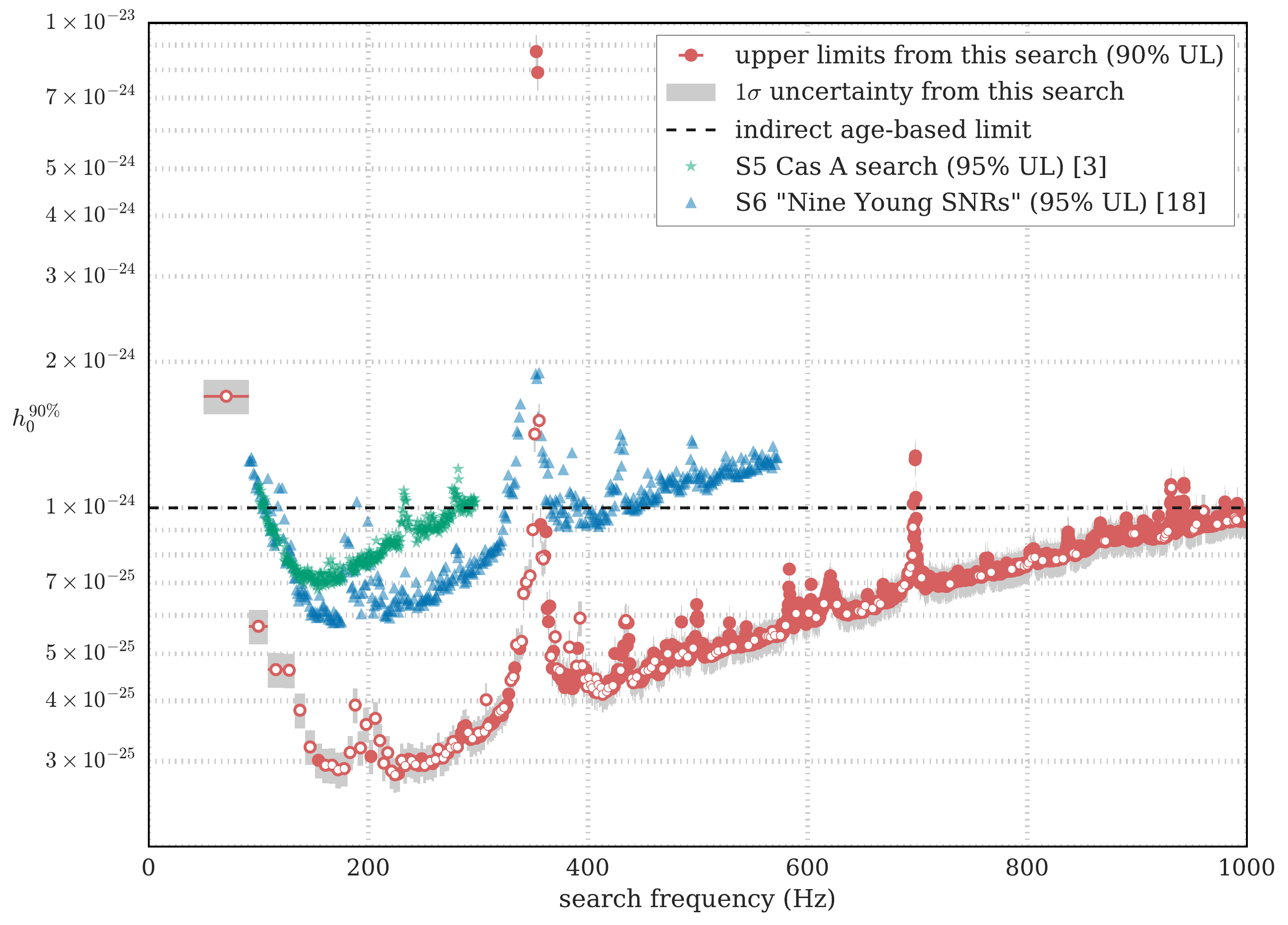}
  \caption{$90\%$ confidence strain amplitude upper limits in each of the
    2000 partitions. The results for partitions that contain only
    undisturbed 50-mHz bands are plotted in the filled red circles, while the
    results for partitions with disturbed 50-mHz bands
    are plotted in the open red circles. We also plot the 95\% confidence upper limits
    from two previous searches on Cas A in green and blue. Our upper limits beat the so-called indirect age-based limit \cite{Wette:2008hg} across the band.}
  \label{upperLimitsPlot}
\end{figure*}

We find no candidates with $CR > 5$ and no excess
in the p-value distribution. Therefore, we set frequentist 90\% upper limits
on the continuous gravitational-wave strain $h_0^{90\%}$ in our search range
using the process described in previous works \cite{S6Bucket,S5GC1HF}, which we summarise below. 

The $h_0^{90\%}$ in a partition is the gravitational-wave amplitude at which
90\% of a population of signals with parameters within the partition would
produce a more significant candidate than the most significant candidate measured
by the search in that partition. We determine $h_0^{90\%}$ by injecting signals
at fixed amplitudes bracketing the $h_0^{90\%}$ level, then running the search on these
injections and counting how many injections were recovered (i.e., how many produced
a candidate more significant
than the loudest measured by the actual search). Because this injection-and-recovery
procedure is time-consuming, we perform it on only a subset of twenty representative
partitions --- uniformly distributed in frequency in the search range --- rather than the
full set of 2000, and use these results to derive the upper limits in all the other partitions.

For each of the twenty injection partitions, we fit a sigmoid to
the detection efficiency (the fraction of recovered injections) as a function of injection amplitude
to determine both the value of $h_0^{90\%}$ and the 1-$\sigma$ uncertainty
on $h_0^{90\%}$. 
We determine the $h_{0,\CR_i}^{90\%,j}$ in each of the injection partitions
corresponding to different detection criteria binned by CR, with
$\CR_i=[0,~ 1,~  2,~  3,~  4,~  5]$. For each $\CR_i$, we derive the corresponding sensitivity depths 
\begin{equation}
\SDd_{\CR_i}^{90\%,j}= {{\sqrt{S_h(f_j)}}\over{h_{0,\CR_i}^{90\%,j}}} ~~~ [1/\sqrt{\text{Hz}}].
\label{eq:SDj}
\end{equation}
By design, the sensitivity depths of this search are roughly constant across the different
partitions. We estimate the sensitivity depths by averaging the values across the injection partitions:
\begin{equation}
\SDd_{\CR_i}^{90\%}={1\over 20} \sum_{j=1}^{20} \SDd_{\CR_i}^{90\%,j}.
\label{eq:SD}
\end{equation}
For each of the remaining partitions, at frequencies around $f_k$, we derive the upper limit as
\begin{equation}
{ h{_{0}^{90\%}}(f_k) } = 
{
{\sqrt{S_h(f_k)}} \over { 
{\SDd}_{\CR_i(f_k)}^{90\%} 
}
},
\label{eq:ULfromSensDepth}
\end{equation}
where $\CR_i(f_k)$ is the significance bin of the loudest candidate of the partition at $f_k$ and
$S_h(f_k)$ the power spectral density of the data. $\SDd_{\CR_i}^{90\%}\simeq70$~Hz$^{-1/2}$ for this search.

Our upper limits are plotted in Fig.~\ref{upperLimitsPlot} in red
with 1-$\sigma$ uncertainties in gray, and provided in tabular form as Supplemental Material.
The uncertainties in $h_0^{90\%}$ that we report here are propagated
from the statistical uncertainties in fitting the recovery.
The partitions containing disturbed
bands (which were not included in the analysis) are marked with open circles.

The upper limit value near 170 Hz, where the detectors are the most sensitive,
is $2.9\times 10^{-25}$. This value is roughly two times lower than the previous most constraining upper limit 
on Cas A \cite{NineYoungSNRs}, plotted in blue, which also used S6 data.
Our upper limits are also more than twice as constraining as
an earlier Cas A search, plotted in green
\cite{S5CasA}, which ran on S5 data\footnote{However, we note that the
other two searches produced 95\% upper limits rather
than 90\% upper limits; the latter is the standard for the
broad surveys by \EatH~ \cite{S6Bucket,S6BucketFUs,S5GC1HF}. The ratio
between the 90\% and the 95\% confidence upper limits is $\sim$1.1.}.
Our upper limits beat the so-called indirect age-based
limit \cite{Wette:2008hg} across the vast majority of the frequency range.

\section{Conclusions}

\begin{figure*}
  \includegraphics[width=0.8\textwidth]{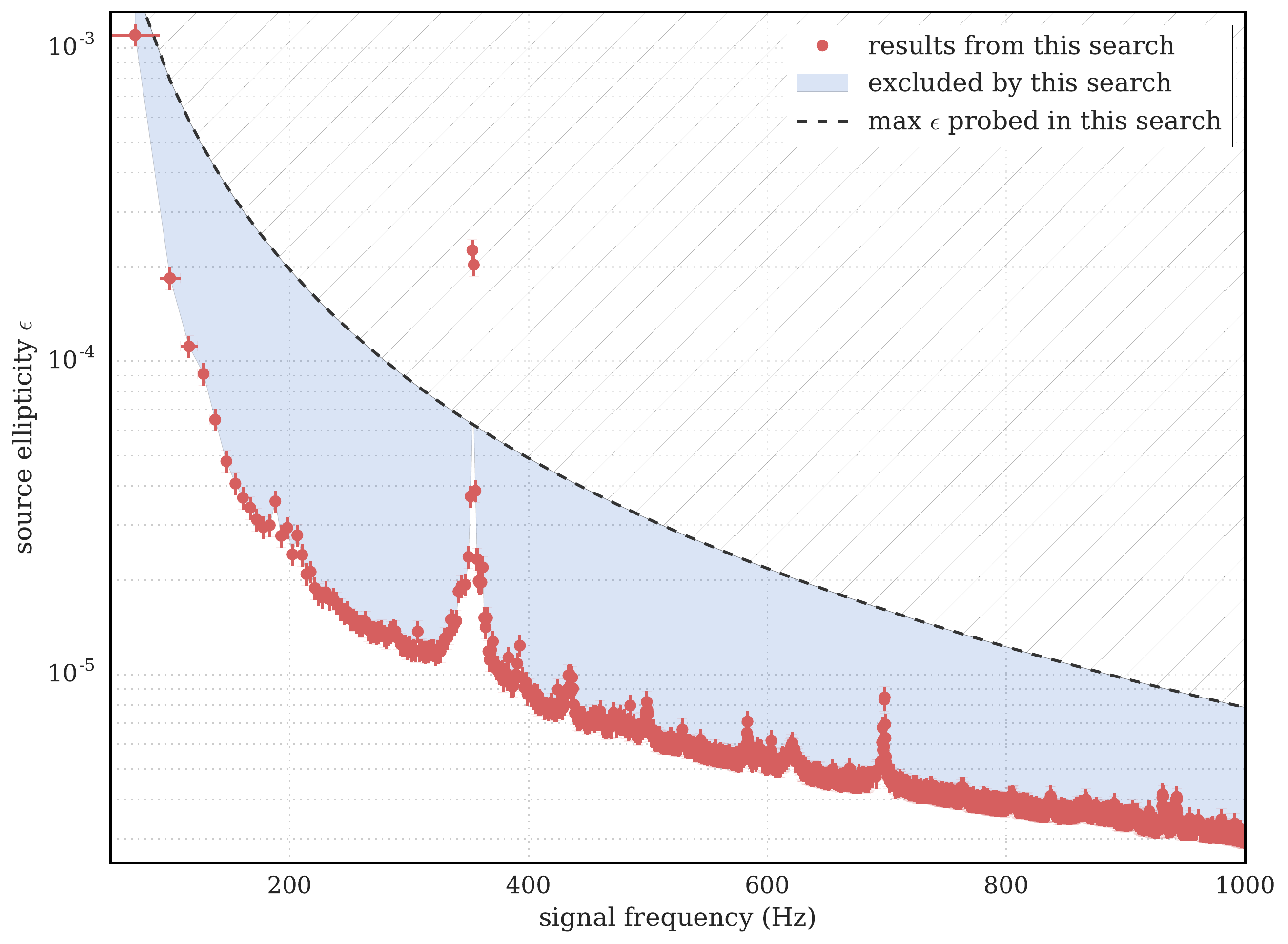}
  \caption{With a few reasonable assumptions, we can convert the 
    upper limits on the gravitational-wave strain to upper limits on
    the ellipticity of Cas A. The shaded area denotes the source ellipticities
    (as a function of signal frequency) that are excluded by this search: 
    ellipticities in this region would have produced signals that this search would have detected. The
    dashed line marks the spindown ellipticity probed by this search and is
    set by our choice of $\dot{f}$ search range.}
  \label{ellipticities}
\end{figure*}

The upper limits on the gravitational wave strain from Cas A
translate into constraints on the shape of Cas A.
As described in \cite{Ming2016}, a neutron star's mass distribution
can be described by the ellipticity $\epsilon$, where
\begin{align}
  \epsilon = \frac{\left|I_{xx} - I_{yy}\right|}{I_{zz}}
\end{align}
and $I_{zz}$ is the principal moment of inertia of the star around
its rotational axis.
If a neutron star at a distance $D$ and spinning at a frequency $f/2$ has a
non-axisymmetric distortion $\epsilon$, then it will produce a continuous
gravitational wave with a frequency $f$ and amplitude $h_0$. These quantities are related to each other as follows:
\begin{align}
  \epsilon = \frac{h_0 D}{f^2} \frac{c^4}{4 \pi^2 I_{zz} G}.
  \label{eq:epsilonVsh0}
\end{align}
Eq.~\ref{eq:epsilonVsh0} shows how we can re-express the constraints
on the gravitational wave amplitude as constraints on the ellipticity.
We take the distance to Cas A to be 3.4~kpc  \cite{CasA:age} and $I_{zz}$
to be $10^{38}$~kg\,m$^2$.

These constraints on source ellipticity are shown in Fig.~\ref{ellipticities}.
For instance, if Cas A is emitting gravitational waves at around 200~Hz
(and, therefore, spinning at a frequency of 100~Hz), its ellipticity should be
less than a few times $10^{-5}$, since we would have been able to detect  
gravitational waves produced by larger ellipticities. 

The maximum ellipticity is the ellipticity necessary to sustain emission at the spindown
limit, i.e., when all of the lost rotational energy is radiated as gravitational waves. This spindown ellipticity is
\begin{equation}
  \epsilon^{sd} = \sqrt{ \frac{5c^5}{32\pi^4G} \frac{x|\dot{f}|}{If^5}} ~~\textrm{with~ x=1}, 
  \label{eq:spindownEpsilon}
\end{equation}
where $\fdot$ is twice the spin-frequency derivative. 

The highest spin-down ellipticity for an object emitting gravitational
waves at a frequency $f$ that our search could have detected can be
computed from Eq.~\ref{eq:spindownEpsilon} by setting $\fdot=f/300$ years.
For an isolated system, if $\fdot$ is twice the spin-frequency derivative,
larger ellipticities would violate energy conservation. For this reason we
only highlight the region between the ellipticity
upper limit curve and the spindown ellipticity curve
as excluded by the search. However, we note that
systems in general could have ellipticities larger than the spindown ellipticity
if the gravitational wave $\fdot$ (the apparent $\fdot$) differs from the
intrinsic one due to, for example, radial motion. 

\section{Acknowledgments}

The authors thank the \EatH ~volunteers who have supported this work by donating compute cycles of their machines.
Maria Alessandra Papa, Bruce Allen and Xavier Siemens gratefully acknowledge the support from $\mathrm{NSF\;PHY}$
Grant 1104902. All the post-processing computational work for this search was
carried out on the $\mathrm{ATLAS}$ super-computing cluster at the
Max-Planck-Institut f{\"u}r Gravitationsphysik / Leibniz Universit{\"a}t
Hannover. We also acknowledge the Continuous Wave Group of the LIGO Scientific
Collaboration for useful discussions.  This document has been assigned LIGO Laboratory
document number \texttt{LIGO-P1600212}.

\newpage

\bibliography{bibtexStyle}




\end{document}